\documentclass[12pt,a4paper]{article}
\usepackage[utf8]{inputenc}
\usepackage[english]{babel}
\usepackage{amsmath}
\usepackage{amsfonts}
\usepackage{amssymb}
\usepackage{graphicx}
\usepackage{authblk}
\usepackage[left=2cm,right=2cm,top=2cm,bottom=2cm]{geometry}
\usepackage{hyperref}
\author[1]{Paul Raux}
\author[2]{Felix Ritort}

\affil[1]{Université Paris Cité, CNRS, LIED, F-75013 Paris, France}
\affil[1]{Université Paris-Saclay, CNRS/IN2P3, IJCLab, 91405 Orsay, France}
\affil[2]{Small Biosystems Lab, Condensed Matter Physics Department, University of Barcelona, Barcelona, Spain and 2 Institut de Nanociència i
Nanotecnologia (IN2UB), Universitat de Barcelona, Barcelona, Spain}

\title{N-States Continuous Maxwell Demon}
\providecommand{\keywords}[1]
{
  \small	
  \textbf{\textit{Keywords---}} #1
}
\begin{document}
\maketitle

\abstract{Maxwell's demon is a famous thought experiment and a paradigm of the thermodynamics of information. 
 It is related to Szilard's engine, a two-state information-to-work conversion device in which the demon performs single measurements and extracts work depending on the state measurement outcome.  A variant of these models, the continuous Maxwell demon (CMD), was recently introduced by Ribezzi-Crivellari and Ritort where work was extracted after multiple repeated measurements every {time}  that $\tau$ is in a two-state system. The CMD was  able to extract unbounded amounts of work at the cost of an unbounded amount of information storage. In this work, we built a generalization of the CMD to the N-state case. We obtained generalized analytical expressions for the average work extracted and the information content. We show that the second law inequality for information-to-work conversion is fulfilled. We illustrate the results for N-states with uniform transition rates and for the $N=3$ case. }\\
 
\keywords{Maxwell demon, Information-to-Work Conversion, Correlated Measurements} 
\section{Introduction}
\label{sec:intro}
\begin{figure}
\centering
\includegraphics[width=12cm]{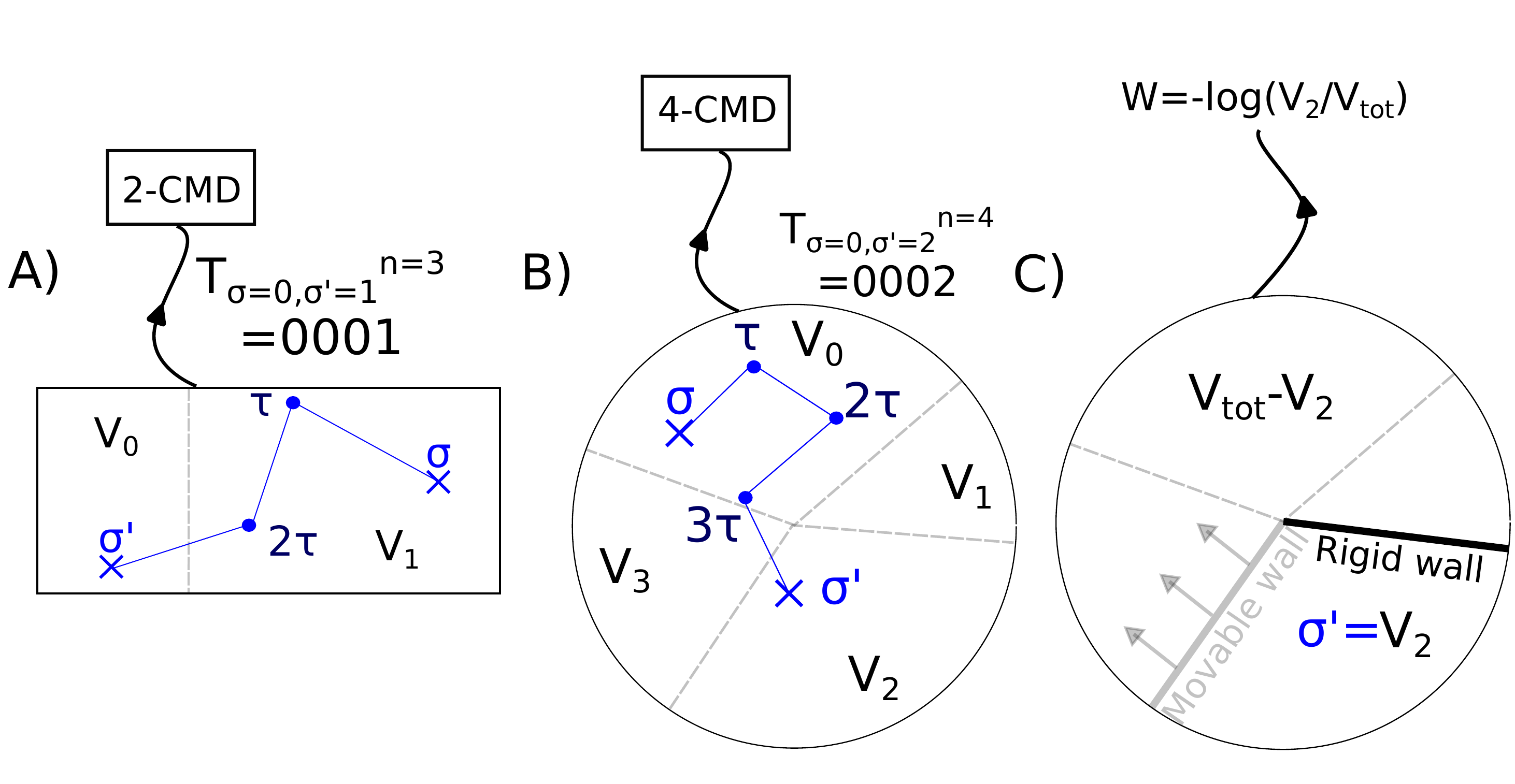}
\caption{
    (\textbf{A}) The 2-CMD   
 is represented as a two-compartment box in which a work extraction protocol is implemented (see text). The cycle of measurement is here $T_{0,1}^3=\underbrace{1,1,1}_{n=3},0$. The average work extracted for this cycle is $-\log \frac{V_1}{V}$. (\textbf{B}) 4-CMD in circular geometry. Each compartment had volume $V_i$. The cycle of measurement of the CMD reads: $T_{0,2}^4=\underbrace{0,0,0,0}_{n=4},2$. The initial state $\sigma$ is $0$ and the final state $\sigma'$ is $2$, the crossing of compartment $3$ remained unnoticed for measurements made at every time $\tau$. (\textbf{C}) In the work extraction protocol, a pair of walls limiting the volume of the last compartment, here $V_2$, are inserted. The wall between compartments $1$ and $2$ is fixed, whereas the wall between compartments $2$ and $3$ was movable and had no mass. To extract the work produced by the expansion of the particle confined in $2$, the movable wall is connected to a pulley device. The average work extracted for this cycle is $-k_BT\log\frac{V_2}{V}$.} 
    \label{fig:figure introduction}
\end{figure}

In 1867, James Clerk Maxwell proposed a thought experiment for the better understanding of the scope and limitations of the second law \cite{leff2002maxwell}. Known as the Maxwell demon paradox, it has spurred  strong research activity for many years, setting the basis for the thermodynamics of information and information-to-work conversion \cite{forgetting,inventions,MDreview,ExpST,Barato2014,Barato2014a,Berut2013,Berut2015,Lutz2015}. In 1929, Leo Szilard introduced a simple physical model \cite{Szilard1929} in which a particle was free to move in a box of volume V with two compartments (denoted with 0 and 1) of volumes $V_0$ and $V_1$ and $V=V_0+V_1$ (Figure~\ref{fig:figure introduction}A). In Szilard's engine (SZ), the ``demon” was an entity able to monitor the particle's position and store the observed compartment in a single-bit variable $\sigma=0,1$.  Information-to-work conversion is as follows: once the particle's compartment is known, a movable wall is inserted between the two compartments, and an isothermal process is implemented to extract work. A work-extracting cycle concludes when the movable wall reaches its far end, and the measurement-work extraction process restarts. The average work extracted per cycle equals the equivalent heat transferred from the isothermal reservoir to the system: $W^{Sz}_{2}=-k_B T(P_{0}\log(P_{0})+P_{1}\log(P_{1}))$, with $P_{0,1} = V_{0,1}/V$ the occupancy probabilities of each compartment. For equal compartments $P_0=P_1=1/2$, Szilard's engine can extract maximal work determined by the Landauer bound, $W^{SZ}\le k_BT\log(2)$ from the reservoir without energy consumption, meaning that heat was  fully converted into work, apparently violating Kelvin's postulate. In the 1960s and 1970s, work by Landauer \cite{Landauer} and Bennett \cite{Bennett1982} found a solution to the paradox. The solution to this paradox considers the information content of the measurement, the work extraction, and the resetting processes of the demon \cite{sagawa2013information,parrondo2015thermodynamics}. Most importantly, to recover the initial state at the end of the thermodynamic cycle, the demon must erase the information acquired on the system~\cite{forgetting}. The minimal erasure cost per bit of information equals $k_BT \log(2)$ for equally probable outcomes. In the end, the information content stored in the demon is always larger than or equal to the extracted work, in agreement with the second law.

In a recent paper, a new variant of the Maxwell demon,  the continuous Maxwell demon (CMD), was introduced \cite{Ribezzi-Crivellari2019nature} (see also \cite{Ribezzi-Crivellari2019JSM} for additional results), analytically solved, and experimentally tested. In the CMD, the demon performs repeated measurements of a particle's location in a two-compartment box every time $\tau$. The first time that the demon measures that the particle  changed compartments, a work extraction procedure is implemented. The main difference with the SZ engine is that, in the CMD, a work extraction cycle contains multiple measurements, whereas for the SZ, a single measurement is performed at every work cycle. 
Compared to the SZ, the CMD can extract a more significant amount of work $W$ because of the larger information content of the multiple stored bits in a cycle. Interestingly, the average work per cycle in the CMD satisfies $W^{CMD}\ge k_BT\log(2)$ being unbounded in the limits $P_0\to 0$ ($P_1\to 1$) and $P_0\to 1$ ($P_1\to 0$). A model combining the SZ and CMD work extraction protocols version showed the role of temporal correlations in optimizing information-to-energy conversion \cite{garrahan2021generalized}. In the CMD, the time between measurements $\tau$ is arbitrary. In particular, it can be made infinitesimal, $\tau\to 0$, leading to an infinite number of measurements per cycle justifying the {\em continuous} adjective of the model.

Here, we generalize the CMD to the case of N-states ($N$-CMD). In a possible realization of the $N$-CMD, a particle in a box of volume $V$ can occupy $N$ distinct compartments of volumes $V_i$ (Figure \ref{fig:figure introduction}B). The demon measures in which compartment the particle is at every time $\tau$ until a change in the compartment is detected. Then, a work extraction process is implemented by inserting one fixed wall at one side and one movable wall at the other side of the compartment that can expand under the elastic collisions exerted by the particle (Figure~\ref{fig:figure introduction}C). A pulley mechanism is attached to the movable wall to extract an average work equal to $W=-k_BT\log P_i$ with $P_i=V_i/V$. For {\emph{N} = 2,}  
 we obtain the standard CMD (Figure~\ref{fig:figure introduction}A) (2-CMD), which corresponds to transforming the Szilard box (Figure~\ref{fig:figure introduction}B) into a periodic torus (Figure~\ref{fig:figure introduction}C).

The outline of this work is as follows. In Section 2, we show how to generalize the mathematical formalism of the 2-CMD to $N$ states ($N$-CMD). In  Section 3, we analyze the performance of the $N$-CMD by studying the thermodynamic efficiency and power. In Section 4, we analyze several cases, and in particular  case $N=3$, to investigate the effect of topology on information-to-work conversion (IWC). We end with some conclusions and future directions. 
\section{General setting}
\label{sec:setting}
Let $\sigma$ ($=1,\ldots,N$) denote the $N$ states of a system following Markovian dynamics defined by transition rates that satisfy detailed balance, ensuring that the system relaxes to an equilibrium state. Let $\tau$ be the time between consecutive measurements. The conditional probability $p_\tau(\sigma'\vert \sigma)$ that the outcome of the measurement is $\sigma'$ after time $\tau$ conditioned that it starts in $\sigma$ satisfies the following master equation:
\begin{equation}
    \partial_\tau  p_\tau (\sigma'\vert \sigma )=\sum_{\sigma''=1}^NK_{\sigma'\sigma''}p_\tau(\sigma''\vert \sigma)\label{eq:master equation}
\end{equation}
with initial condition $p_{\tau\rightarrow 0}(\sigma'\vert \sigma)\rightarrow \delta_{\sigma^{'},\sigma}$, where $\delta$ is the Kronecker delta function. 
 Markov matrix $K_{\sigma'\sigma''}$ satisfies $\sum_{\sigma'}K_{\sigma'\sigma''}=0;\forall\sigma''$, defining transitions rates from state $\sigma''$ to~$\sigma'$:
\begin{equation}
K_{\sigma'\sigma''}=\left\{
\begin{array}{ll}
      -\left(\sum_{\sigma (\neq \sigma'')} k_{\sigma\leftarrow\sigma''}\right) & \; \mathrm{if} \; \sigma'=\sigma'' \\
    k_{\sigma'\leftarrow \sigma''} & \mathrm{otherwise}
\end{array} 
\right. \label{eq:markov matrix}
\end{equation}
with $k_{\sigma'\leftarrow\sigma''}$ the probability to jump from state $\sigma''$ to $\sigma'$ during time $d\tau$.
Let us denote by $P_\sigma$ the stationary solutions of Equation~(\ref{eq:master equation}). The detailed balance condition reads:
\begin{equation}
\forall \sigma,\sigma' \;; K_{\sigma\sigma'}P_{\sigma'}=K_{\sigma'\sigma}P_\sigma
\label{eq:DB}
\end{equation}
The solution of Equation~(\ref{eq:master equation}) can be written using the Perron--Frobenius theorem (see~\cite{Kampen2007}) as  a spectral expansion in terms of the eigenvalues and eigenvectors of $K$:
\begin{equation}
    p_\tau(\sigma'\vert \sigma)=P_\sigma \sum_\alpha l_{\sigma'}^\alpha l_{\sigma}^\alpha \exp (\lambda^\alpha \tau )\label{eq:master equation solution 1}
\end{equation}
where  $l^\alpha$ is the left eigenvector of $K$ associated with the eigenvalue $\lambda^\alpha$. The sum over the $\alpha$ term in Equation~(\ref{eq:master equation solution 1}) is symmetric in $\sigma\leftrightarrow \sigma'$. Therefore, the conditional probabilities also fulfil a detailed balance:
\begin{equation}
    \frac{p_\tau(\sigma \vert \sigma')}{p_\tau(\sigma'\vert \sigma)}=\frac{P_\sigma}{P_{\sigma'}}\label{eq:conditional probability DB}
\end{equation}

\textbf{Remark:} 
Detailed balance ensures that there exists a unique stationary state $P_\sigma$ associated with the eigenvalue $\lambda^0=0$ and that the other eigenvalues are real and negative, $\lambda^{\alpha\ne 0}<0$. Equation (\ref{eq:master equation solution 1}) can be rewritten as follows:
\begin{equation}
    p_\tau(\sigma'\vert \sigma)=P_{\sigma'} \left(1 + \sum_{\alpha \neq 0} l_{\sigma'}^\alpha l_{\sigma}^\alpha \exp (\lambda^\alpha \tau )\right)\label{eq:master equation solution 2}
\end{equation}
which gives $p_\tau(\sigma'\vert \sigma)=P_{\sigma'}$ for $\tau\to\infty$ as expected.

In the CMD, a work-extraction cycle is defined by a sequence of $n+1$ measurement outcomes $\sigma_i$ ($1\le i\le n+1$) repeatedly taken every time $\tau$. In a cycle the first $n$ outcomes are equal ($\sigma$) ending with $\sigma'$ ($\neq\sigma$). We define the trajectory for a cycle as follows: 
\begin{equation}
    T_{\sigma\sigma'}^n=\underbrace{\sigma, \ldots ,\sigma,}_{n} \;\sigma'
    \label{eq:cycle definition}
\end{equation}
The probability of a given trajectory $T_{\sigma\sigma'}^n$ reads:
\begin{equation}
    P\left( T_{\sigma\sigma'}^n \right)= p_\tau (\sigma \vert \sigma)^{n-1}p_\tau(\sigma'\vert \sigma)
    \label{eq:probability of a cycle}
\end{equation}
This is normalized as follows:
\begin{equation}
    \sum_{\sigma'(\neq \sigma)}\sum_{n=1}^\infty P\left( T_{\sigma\sigma'}^n \right)=1\;\;,\forall \sigma
     \label{eq:normalization of cycle probas}
\end{equation}
\subsection{Thermodynamic Work and Information-Content}
\label{subsec:workinfo}
Like in the SZ, the work extracted by the CMD in a given cycle $T_{\sigma\sigma'}^n$ equals $-\log(P_{\sigma'})$. Averaging over all the possible measurement cycles, we obtain the average extracted work:
\begin{equation}
\begin{split}
    W_N(\tau)&=<-\log P_{\sigma'}>=-\sum_{\sigma}\sum_{\sigma'(\neq\sigma)}\sum_{n=1}^{\infty}P_{\sigma}P\left( T_{\sigma\sigma'}^n \right)\log P_{\sigma'}\\
    &=-\sum_{\sigma=1}^N \frac{P_\sigma}{1-p_\tau(\sigma \vert \sigma)}\sum_{\sigma'\neq \sigma}p_\tau (\sigma'\vert \sigma)\log P_{\sigma'}
    \label{eq:definition of the average work extraction}
\end{split}
\end{equation}
which is positive by definition. In the limit $\tau\to\infty$ we obtain the following expression,
\begin{equation}
    W_N^{\infty}=-\sum_{\sigma=1}^N \frac{P_\sigma}{1-P_\sigma}\sum_{\sigma'\neq \sigma}P_{\sigma'}\log P_{\sigma'}\label{eq:definition of the average work extraction for tau infinity},
\end{equation}
which can be written as follows:
\begin{equation}
    W_N^{\infty}=\Bigl(\sum_{\sigma=1}^N \frac{P_\sigma}{1-P_\sigma}\Bigr)W_N^{\rm SZ}+\sum_{\sigma=1}^N \frac{P_\sigma^2\log P_{\sigma}}{1-P_\sigma}
    \label{eq:definition of the average work extraction for tau infinity another expression}
\end{equation}
where $W^{SZ}_N$ is the classical statistical entropy of the system, which can also be interpreted as the average work extraction of the $N$-states Szilard engine, denoted as $N$-SZ:
\begin{equation}
\begin{split}
    W^{SZ}_N=<-\log(P_\sigma)>=-\sum_\sigma P_\sigma \log P_\sigma
\end{split}
    \label{eq:definition of the $N$-SZ work}
\end{equation}

This expression can be readily minimized in the space of $P_{\sigma}$ giving the uniform solution, $P_{\sigma}=1/N$ for which $W_N^{\infty}=\log N$. In contrast, $W_N^{\infty}\to -\log (1-P_\sigma)$ if $P_\sigma\to 1$ for a given $\sigma$ (and $P_{\sigma'}\to 0$ $\forall\sigma'\neq\sigma$) diverging in that limit.   

We define the average information content per cycle as the statistical entropy of the measurement-cycle probabilities \cite{thomas2006elements}:
\begin{equation}
\begin{split}
    I_N(\tau)&=<-\log \left( P_\sigma P\left(T_{\sigma\sigma'}^n \right)\right)>\\
    &=-\sum_{\sigma}\sum_{\sigma'(\neq\sigma)}\sum_{n=1}^{\infty}P_{\sigma}P\left( T_{\sigma\sigma'}^n \right)\log \left(P_\sigma P\left( T_{\sigma\sigma'}^n \right)\right)\\
    &=-\sum_{\sigma=1}^{N}\sum_{\sigma'\neq \sigma}^{N}P_\sigma\left(p_\tau(\sigma'\vert \sigma)\log(P_\sigma)+p_\tau(\sigma'\vert \sigma)\log(p_\tau(\sigma'\vert \sigma))\right)\underbrace{\sum_{n=1}^{\infty}p_\tau (\sigma\vert \sigma)^{n-1}}_{\frac{1}{1-p_\tau(\sigma\vert \sigma)}}\\
    &-\sum_{\sigma=1}^{N}\sum_{\sigma'\neq \sigma}^{N}P_\sigma p_\tau(\sigma'\vert \sigma)\log(p_\tau(\sigma\vert \sigma))\underbrace{\sum_{n=1}^{\infty}(n-1)p_\tau (\sigma\vert \sigma)^{n-1}}_{\frac{p_\tau(\sigma\vert \sigma)}{(1-p_\tau(\sigma\vert \sigma))^2}}\\
    &=W_N^{SZ}-\sum_\sigma^N \frac{P_\sigma}{1-p_\tau(\sigma \vert \sigma)}\sum_{\sigma'}p_\tau (\sigma'\vert \sigma)\log p_\tau (\sigma'\vert \sigma)\label{eq:def average information content}
\end{split}
\end{equation}
The positivity of $I_{N}(\tau)$ follows from the fact that  $p_\tau(\sigma'\vert \sigma),P_\sigma\le 1$. The second term in Equation~(\ref{eq:def average information content}) depends on $\tau$ and can be understood as the contribution of correlations between measurements to $I_{N}(\tau)$. \\
Lastly, using Equation~(\ref{eq:conditional probability DB}), we can rearrange Equation~(\ref{eq:def average information content}) as follows:
\begin{equation}
    I_N(\tau)=W_N(\tau)\underbrace{-\sum_\sigma^N \frac{P_\sigma}{1-p_\tau(\sigma \vert \sigma)}\sum_{\sigma'}p_\tau (\sigma'\vert \sigma)\log p_\tau (\sigma\vert \sigma')}_{=\Delta_N> 0}\label{eq:second law form of information}
\end{equation}
where the second term $\Delta_N$ is positive since $p_\tau (\sigma \vert \sigma')\le 1$. Equation~(\ref{eq:second law form of information}) implies  the second law inequality:
\begin{equation}
    I_N(\tau)-W_N(\tau)>0 \; \forall \tau 
    \label{eq:second law inequality}
\end{equation}
meaning that the cost to erase the stored sequences information content is always larger than the work extracted by the demon.

\subsection{Comparison with the Szilard Engine}
\label{subsec:comparison with SZ}
Equating Expressions (\ref{eq:def average information content}) and (\ref{eq:second law form of information}) for $I_N(\tau)$, we obtain a relation between $W_N(\tau)$ and $W^{SZ}$ that compares the average work extracted in the $N$-CMD to the $N$-SZ engine as~follows:
\begin{equation}
\begin{split}
    W_N(\tau)-W_N^{SZ}&=<-\log \frac{P_{\sigma'}}{P_\sigma} > \\
    &=-\sum_\sigma \frac{p_\sigma}{1-p_\tau(\sigma \vert \sigma)}\sum_{\sigma'}p_\tau (\sigma' \vert \sigma) \log \frac{p_\tau (\sigma' \vert \sigma)}{p_\tau (\sigma \vert \sigma')}\ge 0
    \label{eq:comparison with the Szilard}
    \end{split}
\end{equation}
where the first equality follows from the difference between the first right-hand side of Equations~(\ref{eq:definition of the average work extraction}) and (\ref{eq:definition of the $N$-SZ work}).
This shows that the CMD's average work per cycle is always larger or equal to SZ. The equality holds for the uniform case $P_{\sigma}=1/N$ where $W_N(\tau)=W_N^{SZ}=\log N$.
\section{Thermodynamic power and efficiency}
\label{sec:thermodynamic efficiency and power}
\subsection{Average Cycle Length}
\label{subsec:LC}
As a preliminary, we first compute the average time of a cycle of measurement. This is similar to the mean first residence time of the system, except for the fact that (unobserved) hopping events are permitted. We define it as follows:
\begin{equation}
    t_N^c\equiv\tau <n>
    \label{eq:def tc}
\end{equation}
and obtain the following expression:
\begin{equation}
    t_N^c=\tau \left( 1 + \sum_\sigma \frac{P_\sigma}{1-p_\tau(\sigma \vert \sigma)}\right)
    \label{eq:expression of tc}
\end{equation}
The following equalities are shown:
\begin{equation}
\lim_{\tau\to 0^+}t_N^c=-\sum_i \frac{1}{\sum_{\alpha\neq 0} (l_i^\alpha)^2\lambda^\alpha}>0
\label{eq:small time limit of tc}
\end{equation}
\begin{equation}
\lim_{\tau\to \infty}t_N^c= +\infty
\label{eq:long time limit of tc}
\end{equation}
The average cycle time is the mean first passage time \cite{benichou2015mean} of the discrete time random walk defined by a cycle of measurements. 
\subsection{Thermodynamic Power}
\label{subsec:power}
We define the thermodynamic power as the average work $W_N$ extracted per cycle time~$t_N^c$:
\begin{equation}
    \Phi_N(\tau)=\frac{W_N}{t_N^c}
    \label{eq:definition of phi}
\end{equation}
In the limit of uncorrelated measurements $\tau\rightarrow \infty$, we obtain from Equations~(\ref{eq:definition of the average work extraction for tau infinity}) and (\ref{eq:expression of tc}):
\begin{equation}
    \Phi_N^{\infty}=-\frac{1}{\tau}\frac{\sum_{\sigma=1}^N \frac{P_\sigma}{1-P_\sigma}\sum_{\sigma'\neq \sigma}P_{\sigma'}\log P_{\sigma'}}{1 + \sum_\sigma \frac{P_\sigma}{1-P_\sigma}}
    \label{eq:phi_infty}
\end{equation}
For $N=2$, we recover the results in  \cite{Ribezzi-Crivellari2019nature,Ribezzi-Crivellari2019JSM}.
\subsection{Information-to-Work Efficiency}
\label{subsec:efficiency}
In the spirit of the efficiencies defined for thermal machines, we define the information-to-work conversion (IWC) efficiency of the CMD as the ratio between $W_N$, taken to be the objective function, and $I_N$, taken to be the cost function, for the optimization of the CMD:
\begin{equation}
    \eta_N=\frac{W_N}{I_N}
    \label{eq:definition of efficiency}
\end{equation}
Using Equation~(\ref{eq:second law form of information}), we can rewrite $\eta_N$ as follows:
\begin{equation}
\eta_N=\frac{1}{1+\frac{\Delta_N}{W_N}}
\label{eq:second law form of efficiency}
\end{equation}
From Equation~(\ref{eq:second law inequality}), $\eta_N<1$. In the limit $\tau\rightarrow \infty$, we obtain:
\begin{equation}
\lim_{\tau\to \infty}\eta_N=\frac{1}{1+\frac{\sum_i \frac{P_i}{1-P_i}\log P_i}{\sum_i \frac{P_i}{1-P_i}\sum_{j\neq i}P_j \log P_j}}
\label{eq:long time limit of efficiency}
\end{equation}
In  limit $P_\sigma\rightarrow 1$ for a given state $\sigma$, one can check that the $N$-CMD reaches  maximal efficiency 1. 
%However, in the context of the optimization of IWC, we have thus schown that efficiency 1 is reached only for vanishing power (see section \ref{sec:thermodynamic efficiency and power}). 
%
\section{Particular cases}
\label{sec:cases}
\section{Particular Cases}
\label{sec:cases}
Here, we analyze some specific examples.
\subsection{Case N=2}
We now turn to the  $N=2$ case considered in \cite{Ribezzi-Crivellari2019nature} as an example of our formulae. The kinetic rate matrix in this case reads:
\begin{equation}
    K=\begin{pmatrix}
        -k_{1\leftarrow 0} & k_{0\leftarrow 1} \\
        k_{1\leftarrow 0} & -k_{0\leftarrow 1}
    \end{pmatrix}
    \label{eq:markov matrix 2-CMD}
\end{equation}
Here, we do not need to make any particular choice of rates $k_{\sigma^{'}\sigma}$ to ensure detailed balance since, for two states, a detailed balance unconditionally holds. Applying the procedure sketched in Section \ref{sec:setting}, we solve the master equation:
\begin{equation}
p_\tau=(p_\tau(\sigma \vert \sigma^{'}))_{\sigma,\sigma^{'}=0,1}
\begin{pmatrix}
    P_0+P_1\exp(-R\tau) & P_0(1-\exp(-R\tau))\\
    P_1(1-\exp(-R\tau)) & P_1 + P_0\exp(-R\tau)
\end{pmatrix}\label{eq:solution master equation 2-CMD}
\end{equation}
where $R=k_{1\leftarrow 0}+k_{0\leftarrow 1}$, $P_0=\frac{k_{0\leftarrow 1}}{R}$ and $P_1=\frac{k_{1\leftarrow 0}}{R}$ such that $P_0+P_1=1$. $p_\tau$ is normalized per column: 
\begin{equation}
p_\tau(\sigma^{'}\vert \sigma)+p_\tau(1-\sigma^{'}\vert \sigma)=1 \;,\;\;\forall \sigma=0,1 \label{eq:normalization 2-CMD}
\end{equation}
First, let us consider $W_2$. Since $N=2$ and by normalization, there is only one term in the sum $\sum_{\sigma^{'}\neq \sigma}$ of Equation~(\ref{eq:definition of the average work extraction}). Thus, $W_2$ simplifies to:
\begin{equation}
    W_2=-P_0\log(1-P_0)-(1-P_0)\log(P_0)
    \label{eq:work 2-CMD}
\end{equation}
We recover the result obtained in \cite{Ribezzi-Crivellari2019nature} and  coincidently show that the $\tau$ independence of this result is a particular feature of the  $N=2$ case. Moreover, since $W_2$ had a simple expression, we obtained a tractable expression for the comparison with the SZ average work extracted, c.f. Equation~(\ref{eq:comparison with the Szilard}):
\begin{equation}
    W_2-W_2^{SZ}=(1-2P_0)\log\bigl(\frac{1-P_0}{P_0}\bigr)
    \label{eq:work comparison 2-CMD}
\end{equation}
This quantity is positive and vanishes only for uniform probability, $P_\sigma=\frac{1}{2}$, as shown in Section~\ref{subsec:LC}. Using  normalization Equation~(\ref{eq:normalization 2-CMD}) again in the definition of $I_N$ Equation~(\ref{eq:def average information content}), we obtain $I_2$ as follows:
\begin{equation}
\begin{split}
    I_2&=-P_0\log P_0-(1-P_0)\log(1-P_0)\\
    &-P_0\left( \frac{p_\tau(0\vert 0)}{p_\tau(1\vert 0)}\log p_\tau(0\vert 0)+\log p_\tau(1\vert 0) \right) \\
    &-(1-P_0)\left( \frac{p_\tau(1\vert 1)}{p_\tau(0\vert 1)}\log p_\tau(1\vert 1)+\log p_\tau(0\vert 1) \right)
\end{split}
    \label{eq:information 2-CMD}
\end{equation}
which is the result obtained in \cite{Ribezzi-Crivellari2019nature}. The remaining results of \cite{Ribezzi-Crivellari2019nature} are obtained by combining Equations (\ref{eq:solution master equation 2-CMD}), \eqref{eq:work 2-CMD}, and \eqref{eq:information 2-CMD}. 

\subsection{Uniform Transition Rates}
\label{subsec:uniform rates}
In this subsection, we take the following particular case for the Markov matrix $K$:
\begin{equation}
K_{\sigma'\sigma}=R \times
    \left\{
\begin{array}{ll}
      -(N-1) & \; \mathrm{if} \;\sigma'=\sigma \\
    1 & \mathrm{otherwise}
\end{array} 
\right.
\label{eq:markov uniform CMD}
\end{equation}
In this case, there are only two independent conditional probabilities; we can thus rewrite the master equation as follows:
\begin{equation}
    \partial_\tau p_\tau(\sigma \vert \sigma)=R(1-N p_\tau(\sigma \vert \sigma))
    \label{eq:Master equation uniform rates}
\end{equation}
Via normalization, we obtain $p_\tau (\sigma'\vert \sigma)$ as follows:
\begin{equation}
    p_\tau(\sigma'\vert \sigma)=\frac{1}{N-1}(1-p_\tau(\sigma\vert \sigma))\,\,\,;\,\,\,\sigma'\neq\sigma
    \label{eq:normalisation uniform}
\end{equation}
In the remainder of this subsection, we  define the dimensionless rescaled time between two measurements as $\Tilde{\tau}=R\tau$.
The solution of Equation~(\ref{eq:Master equation uniform rates}) reads:
\begin{equation}
    p_\tau(\sigma \vert \sigma)=\frac{1}{N}\left(1+(N-1)\exp(-N\Tilde{\tau})\right)
    \label{eq:solution master equation uniform CMD}
\end{equation}
This particular case allows for us to obtain a glimpse of the dependence of the quantities introduced in Section \ref{sec:setting} with N. The average work extracted is as follows:
\begin{equation}
    W_N=\log N\,\,\,\,.
    \label{eq:work uniform CMD}
\end{equation}
We see that the work extracted does not depend on $\tau$. $I_N$ reads:
\begin{equation}
    I_N=\log N-\frac{N}{N-1}\log\bigl(\frac{1}{N}(1-\exp(-N\Tilde{\tau}))\bigr)
    \label{eq:information uniform CMD}
\end{equation}
The first remark is that in the limit $\tilde{\tau} \rightarrow \infty$, $I_N^\infty=\frac{2N-1}{N-1}\underbrace{\log N}_{W_N}$.  \\
One way to optimize the CMD is to maximize  IWC efficiency, defined as follows:
\begin{equation}
    \eta_N\equiv\frac{W_N}{I_N}=\frac{\log N}{\log N-\frac{N \log \left(\frac{1-e^{-N
   \tilde{\tau}}}{N}\right)}{N-1}}
   \label{eq:efficiency uniform CMD}
\end{equation}
We find the asymptotic efficiency $\eta_N^{\infty}=\frac{N-1}{2N-1}$ for $\Tilde{\tau}\rightarrow \infty$ and $\eta_N=\frac{1}{2}$ for $N\rightarrow \infty$. For the thermodynamic power, we obtain:
\begin{equation}
    \Phi_N\equiv\frac{W_N}{t_N^c}=\frac{\log N}{\frac{N \tilde{\tau} e^{N \tilde{\tau}}}{(N-1) \left(e^{N
   \tilde{\tau}}-1\right)}+\tilde{\tau}} 
   \label{eq:power uniform CMD}
\end{equation}
where $t_N^c$ is the average cycle time that we analyzed in Section \ref{subsec:LC}. One can show that the maximum thermodynamic power $\Phi_N=(N-1)\log(N)$ is obtained in the limit $\tilde{\tau}\rightarrow 0$. This shows that the maximum IWC efficiency Equation~(\ref{eq:efficiency uniform CMD}) and the efficiency at maximum power Equation~(\ref{eq:power uniform CMD}) are obtained in two different limits, a general result expected for thermodynamic machines \cite{Goupil2019}.

\subsection{Case N = 3}
\label{subsec:N=3}
\begin{figure}[h!]
\centering
   \includegraphics[width=12cm]{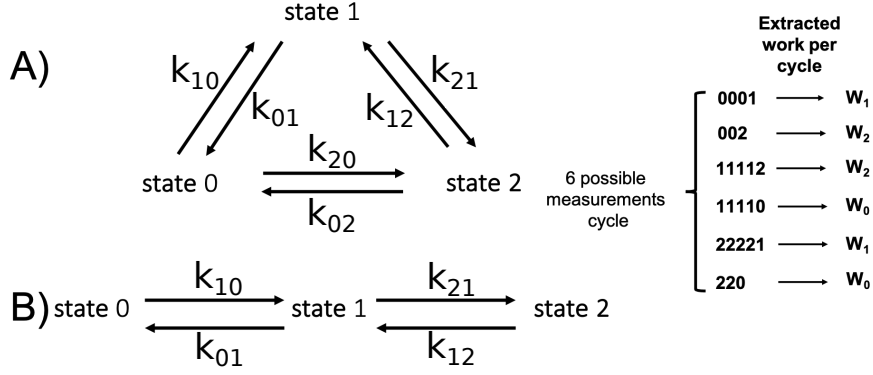}
   \caption{{Definition}  
 of the state spaces for the 2 topologies available for the 3-CMD: (\textbf{A}) Triangular 3-CMD, (\textbf{B}) Linear 3-CMD.}
   \label{fig:3-states topology}
\end{figure}

The 3-CMD is the simplest case in which two different topologies of the state space are available. They are defined in Figure~\ref{fig:3-states topology} and are  denoted   as triangular (Panel A) and linear (Panel B), respectively. We denote the energy of state $\sigma(\sigma=0,1,2)$ by $\epsilon_\sigma$. Taking $\beta=1$, the detailed balance assumption Equation~(\ref{eq:DB}) then reads:
\begin{equation}
    \frac{K_{\sigma\sigma'}}{K_{\sigma'\sigma}}=\exp(- (\epsilon_\sigma-\epsilon_{\sigma'})) \;\;\;,\;\; \forall \sigma\neq\sigma'
    \label{eq:DB function of energies}
\end{equation}
Here we take $\epsilon_0=0$. This implies that the energies of states $1,2$ read:

\begin{equation}
    \epsilon_1=\log(\frac{P_0}{P_1})
    \label{eq:epsilon1}
\end{equation}
\begin{equation}
    \epsilon_2=\log(\frac{P_0}{P_2})
    \label{eq:epsilon2}
\end{equation}
In the linear case, taking as a particular case $k_{01}=1$ and $k_{21}=1$, we obtain the following Markov matrix:
\begin{equation}
        K_3^{lin}=
    \begin{pmatrix}
        -1 & \exp(\epsilon_1) & 0 \\
        1 & -(\exp(\epsilon_1) + \exp(\epsilon_1-\epsilon_2)) & 1\\
        0 & \exp(\epsilon_1-\epsilon_2) & -1
    \end{pmatrix}
    =
    \begin{pmatrix}
        -1 & \frac{P_0}{P_1} & 0 \\
        1 & -\frac{P_0}{P_1}(1+\frac{P_2}{P_0}) & 1\\
        0 & \frac{P_2}{P_1} & -1
    \end{pmatrix}
    \label{eq:linear markov matrix}
\end{equation}
where we  used Equations~(\ref{eq:epsilon1}) and (\ref{eq:epsilon2}) to give an expression of $K_3^{lin}$ depending only on $P_0,P_1,P_2$.
In the triangular case, taking $k_{01}=1$ and $k_{21}=1$ and $k_{02}=1$ as a particular case, we obtain similarly the following Markov matrix:

\begin{equation}
    K_3^{tri}=
    \begin{pmatrix}
        -2 & \exp(\epsilon_1) & \exp(\epsilon_2) \\
        1 & -(\exp(\epsilon_1) + \exp(\epsilon_1-\epsilon_2)) & 1\\
        1 & \exp(\epsilon_1-\epsilon_2) & -(1+\exp(\epsilon_2))
    \end{pmatrix}
    =
    \begin{pmatrix}
        -2 & \frac{P_0}{P_1} & \frac{P_0}{P_2} \\
        1 & -\frac{P_0}{P_1}(1+\frac{P_2}{P_0}) & 1\\
        1 & \frac{P_2}{P_1} & -(1+\frac{P_0}{P_2})
    \end{pmatrix}
    \label{eq:triangular markov matrix}
\end{equation}

The solution of Equation~(\ref{eq:master equation}) with Markov matrix (\ref{eq:linear markov matrix}) in the linear case and (\ref{eq:triangular markov matrix}) in the triangular case, can be written using the Perron-Frobenius theorem \cite{Kampen2007} as the following spectral expansion:
\begin{equation}
    P_\sigma(\tau)=\Psi_0 + c_1^\sigma \Psi_1 \exp(\lambda_1\tau) + c_2^\sigma \Psi_2 \exp(\lambda_2\tau)
    \label{eq:sol master eq N=3}
\end{equation}
where $\lambda_1,\lambda_2<0$ and
%we have denoted $(P_\sigma(\tau))_{\sigma'}=p_\tau(\sigma' \vert \sigma)$ and
$c_1^\sigma,c_2^\sigma$ are the coefficients determined in the limit $\tau\rightarrow 0$, which depend on the conditional state $\sigma$. These coefficients are gathered in table \ref{tabular:coefficients of the spectral expansion} for both models. 

\begin{center}
\begin{tabular}{|c|c|c|}
\hline 
 & $c_1^\sigma$ & $c_2^\sigma$ \\ 
\hline 
$\sigma=0$ & $-(P_2 + \frac{P_2^2+P_2 (P_0+P_1)}{P_0+P_2})$ & $\frac{P_1P_2}{P_0+P_2}$ \\ 
\hline 
$\sigma=1$ & 0 & $\frac{P_2(P_1-1)}{P_0+P_2}$ \\ 
\hline 
$\sigma=2$ & $\frac{P_0}{P_0+P_2}$ & $\frac{P_1P_2}{P_0+P_2}$ \\ 
\hline 
\end{tabular}
\label{tabular:coefficients of the spectral expansion}
\end{center}
\newpage

$\Psi_0,\Psi_1,\Psi_2$ are the eigenvectors of both $K_3^{lin},K_3^{tri}$:
\begin{itemize}
    \item $\Psi_0$ is the eigenvector associated to the eigenvalue $0$ and it corresponds to the stationary probability. Since the detailed balance condition Equation~(\ref{eq:DB}) holds, the stationary probability is the Boltzmann distribution. Thus, 
    \begin{equation}
      \Psi_0=\begin{pmatrix}
        P_0 \\
        P_1 \\
        P_2
    \end{pmatrix}=
    \frac{1}{Z}
    \begin{pmatrix}
        1 \\
        \exp(-\epsilon_1)\\
        \exp(-\epsilon_2)
    \end{pmatrix}
    \label{eq:boltzmann distribution N=3}
    \end{equation}
\noindent where $Z=1+\exp(-\epsilon_1)+\exp(-\epsilon_2)$
    \item $\Psi_1$ is the eigenvector associated to the second eigenvalue, which reads $\lambda_1^{lin}=-1$ in the linear case, and $\lambda_1^{tri}=-(1+\frac{1-P_1}{P_2})$ in the triangular case. $\Psi_1$ reads:
    \begin{equation}
        \Psi_1=\begin{pmatrix}
            -1 \\
            0 \\
            1
        \end{pmatrix}
    \end{equation}
    \item $\Psi_2$ is the eigenvector associated in both models to the eigenvalue $\lambda_2=-\frac{1}{P_1}$. It reads:
    \begin{equation}
        \Psi_2=
        \begin{pmatrix}
            \frac{P_0}{P_2}\\
            -\frac{1-P_1}{P_2}\\
            1
        \end{pmatrix}
    \end{equation}
\end{itemize}

\subsubsection{Uncorrelated Measurements on the 3-CMD}
\label{subsubsec:uncorrelated measurements N=3}
\begin{figure}[h!]
\centering 
    \centering
    \includegraphics[width=18cm]{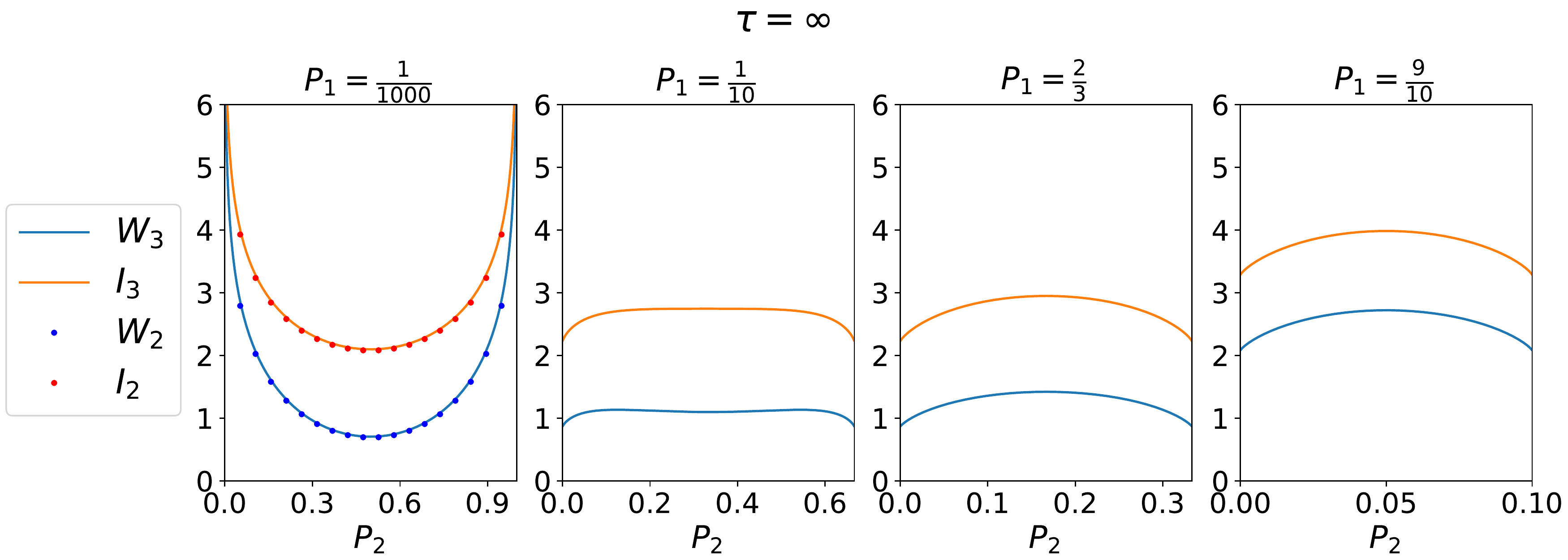}
    \caption{$W_2,I_2,W_3,I_3$ as a function of $P_2$ for $P_1$ fixed in each panel. Large work extraction is obtained in the limit of rare events $P_1 \rightarrow 0$ and $P_2 \rightarrow 0,1$.}
    \label{fig:I W tau infinity}
\end{figure}

We now turn to the limit $\tau\rightarrow \infty$.  In this limit of uncorrelated measurements, the time between consecutive measurements $\tau$ is  larger than the relaxation time of the system, the inverse of the lowest eigenvalue, $\sim -1/\lambda_1$.  In this limit, $P_\sigma(\tau)$ reduces to  Boltzmann distribution Equation~(\ref{eq:boltzmann distribution N=3}) and $p_\tau(\sigma'\vert \sigma)=P_{\sigma'}$. Therefore, the two models (linear and triangular) are indistinguishable. 
Results for work and information are shown in Figure~\ref{fig:I W tau infinity}.
\newpage
 First, it is clear that the second law Inequality (\ref{eq:second law inequality}) was satisfied.  In the limit $P_1\rightarrow 0$, we recovered the 2-CMD. Our generalized expressions for work and information content reproduced well the trend observed in {Figure 1c}  
 of~\cite{Ribezzi-Crivellari2019nature}. In the limit of rare events, where $P_1\rightarrow 0$ and $P_2 \rightarrow 0,1$, we recovered the infinite average work extraction described for the 2-CMD. Large work extraction was only obtained in the 2-CMD limit.\\

\begin{figure}[h!]
\centering %% If there is a figure in wide page, please release command \centering
    \includegraphics[width=18cm]{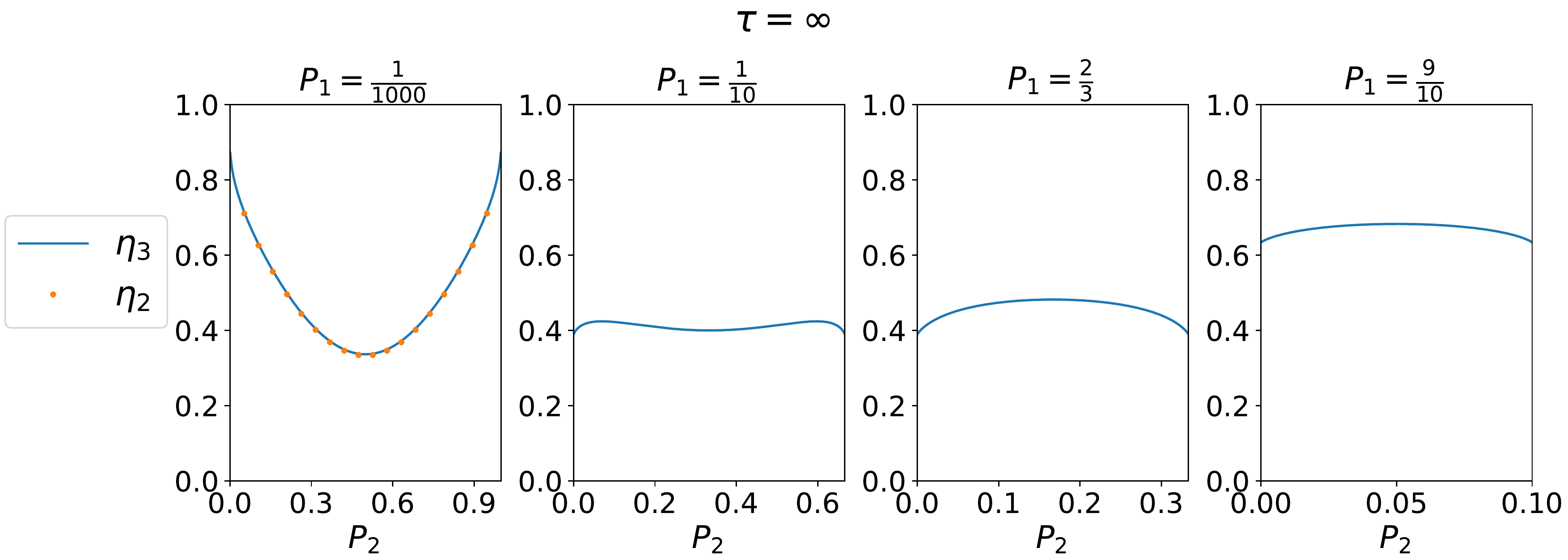}
    \caption{$W_2,I_2,W_3,I_3$ as a function of $P_2$ for $P_1$ fixed in each panel.}
    \label{fig:efficiency tau infinity}
\end{figure}

Efficiency $\eta_3$ is shown in Figure~\ref{fig:efficiency tau infinity}. For $P_1\to 0$ and $P_2\to 0$ or $P_2\to 1$, we recovered the limit of rare events and maximal efficiency $\eta_3\to 1$. In the 3-CMD, we have $\eta_3 \in \left[ 2/5,1 \right]$.
\subsection{Correlated Measurements in the 3-CMD}
\label{subsubsec:correlated meas on 3-CMD}
\begin{figure}[h!]
\centering 
    \centering
    \includegraphics[width=17cm]{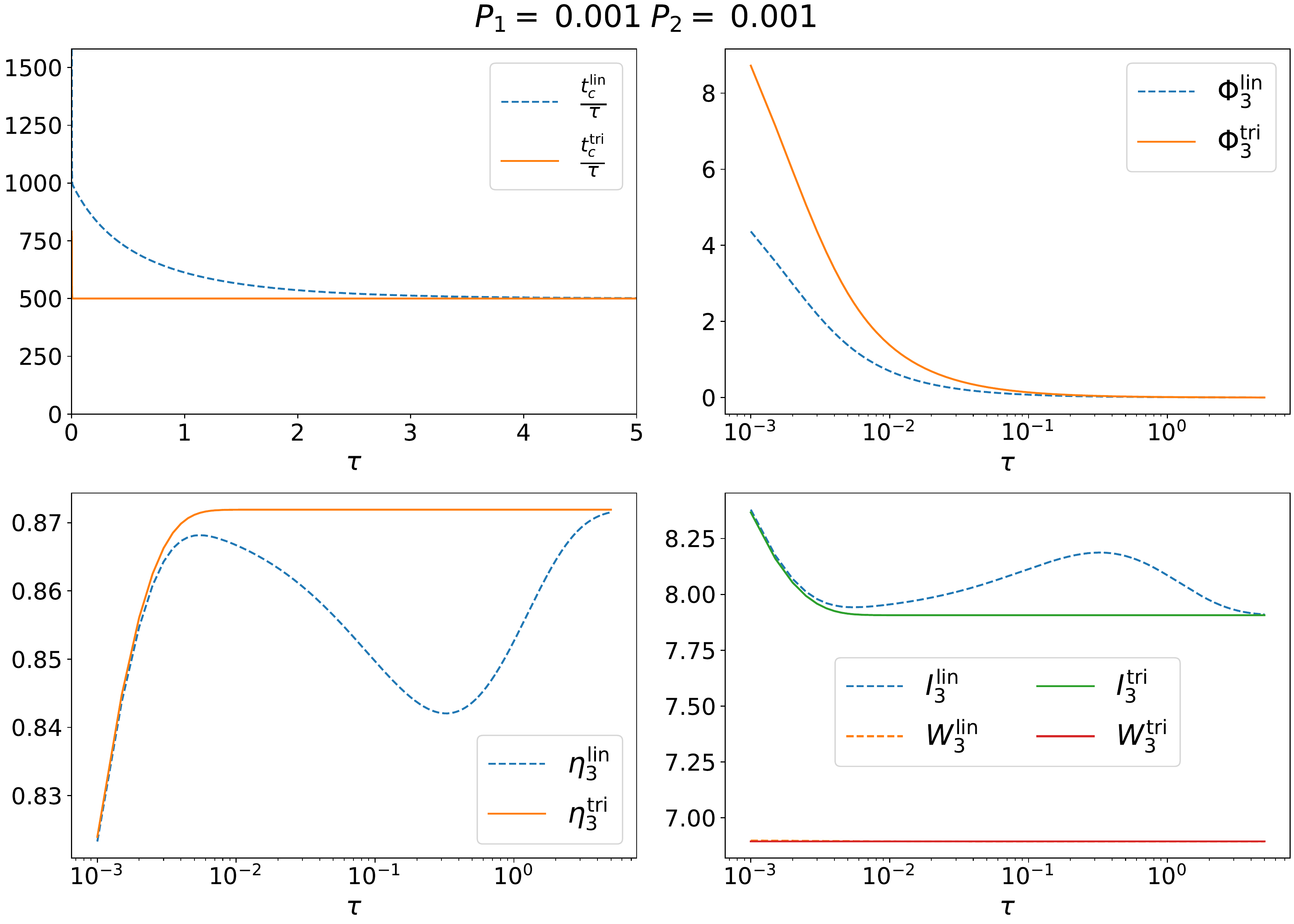}
    \caption{The 3-CMD for correlated measurements for $P_1=P_2=0.001$. (upper left) Average cycle length $t^c_3/\tau$ in both models, Equation~(\ref{eq:expression of tc}); (upper right) thermodynamic power; (lower left) efficiency; (lower right) average information content and work extraction in $k_BT$ units (orange and red lines collapse on top of each other).}
    \label{fig:finite time 3-CMD}
\end{figure}

Correlated measurements are those where $\tau$ is lower than or comparable to the equilibrium relaxation time. Equation~(\ref{eq:sol master eq N=3}) shows that the dynamics of the linear and triangular topologies for the 3-CMD are very similar. Indeed, in the limit of uncorrelated measurements, the two dynamics reduce to the same Boltzmann distribution. They also collapse in the limit $P_1 \rightarrow 1$ (with $P_0,P_2\to 0$), indeed in this case $\lambda_1^{lin}=\lambda_1^{tri}$. In between, the topology of the network is relevant. For correlated measurements, we obtained the results shown in Figure~\ref{fig:finite time 3-CMD}.
First,  the average cycle time (upper-left panel in Figure~\ref{fig:finite time 3-CMD}) in the linear case was  generally larger than that in the triangular case. The direct consequence, since the average work extraction was comparable in both cases, was that the thermodynamic power (upper-right panel in Figure~\ref{fig:finite time 3-CMD})  extracted by the linear 3-CMD was lower than the thermodynamic power extracted by the triangular 3-CMD. Moreover, the thermodynamic power decreased logarithmically to 0 when $\tau$ increases. %The time scale of relaxation to zero power in both cases is of the order of the unity in our system of units. 
Thus,  3-CMD had optimal power production in  limit $\tau \rightarrow 0$, i.e., in the limit of continuous measurements. 
The efficiency of the 3-CMD as a function of $\tau$ is plotted in the lower-left panel of Figure~\ref{fig:finite time 3-CMD}. The linear 3-CMD was  generally less efficient than the triangular 3-CMD. The reason for this is in the lower-right panel of Figure~\ref{fig:finite time 3-CMD}, where $W_3$ and $I_3$ are plotted against $\tau$ for both models. For a comparable work extraction, the linear 3-CMD needs to store more information. Again, in the limit of uncorrelated measurements, the two models converge to the same result. 

\newpage
\section{Concluding Remarks}
\label{sec:concluding remarks}

In this work, we generalized the 2-CMD of~\cite{Ribezzi-Crivellari2019nature,Ribezzi-Crivellari2019JSM} to N-states. We  obtained generalized expressions of the average extracted work, the average information content stored in the demon's memory, and of thermodynamical quantities such as the thermodynamic power and the information-to-work efficiency of the $N$-CMD. We proved that the second law inequality holds for the $N$-CMD, thus giving bounds on the efficiency of the engine. Comparing the $N$-CMD to the $N$-SZ engine, we also showed that the $N$-CMD could extract more work on average than the $N$-SZ engine. The most efficient setting of the $N$-CMD was in the limit of rare events already described in~\cite{Ribezzi-Crivellari2019nature}. In the $N$-CMD case, this limit was obtained by first taking the 2-CMD limit. Thus, no configuration is more efficient in the $N$-CMD than the 2-CMD limit. 

In future work on the $N$-CMD, it would be interesting to implement a graph theoretic procedure to obtain, for instance, a more precise explanation of the difference between the linear and  triangular cases (connected graph versus fully connected graph). It would also be interesting to determine the distributions of the quantities computed here \cite{Broeck2015} and thus optimize the fluctuations of the $N$-CMD. 
\paragraph{Author contributions:} F. R. conceived the work, and P. R. did the calculations.
\paragraph{Acknowledgements:}
FR is supported by the Spanish Research Council Grant  PID2019-111148GB-100 and the Icrea Academia Prize 2018.

\newpage
\bibliographystyle{unsrt}
\bibliography{biblioPubli.bib}

\end{document}